\documentclass[nofootinbib,
twocolumn,
superscriptaddress,
 amsmath,amssymb,
 aps,prc
]{revtex4-2}

\usepackage{amssymb}  
\usepackage{graphicx}
\usepackage{exscale}
\usepackage{textcomp}
\usepackage{enumerate}
\usepackage{amsmath}
\usepackage{amssymb}
\usepackage{bm}
\usepackage{multirow}
\usepackage{orcidlink}
\usepackage{appendix}
\usepackage{slashed}
\usepackage{soul}

\usepackage{color}
\usepackage{gensymb}
\usepackage{dsfont}
\usepackage[normalem]{ulem}  

\newcommand{\De}{\Delta}

\newcommand{\ka}{\kappa}
\newcommand{\la}{\lambda}

\newcommand{\bea}{\begin{eqnarray}}
\newcommand{\eea}{\end{eqnarray}}
\newcommand{\ba}[1]{\begin{array}{#1}}
\newcommand{\ea}{\end{array}}

\newcommand{\beq}{\begin{equation}}
\newcommand{\eeq}{\end{equation}}

\newcommand{\sliver}{\kern 0.07em} 

\newcommand{\fm}{\text{fm}}

\newcommand{\MeV}{\text{MeV}}

\newcommand{\nsat}{\ensuremath{n_{\text{sat}}}}

\begin{document}

\title{Neutrino absorption in two-flavor color-superconducting quark matter}

 \author{M.~G.~Alford \,\orcidlink{0000-0001-9675-7005}
}
 \email{alford@physics.wustl.edu}
 \affiliation{Department of Physics, Washington University in St.~Louis, St.~Louis, MO 63130, USA}
 
 \author{L.~Brodie\,\orcidlink{0000-0001-7708-2073
}}
 \email{b.liam@wustl.edu}
 \affiliation{Department of Physics, Washington University in St.~Louis, St.~Louis, MO 63130, USA}

 \author{M.~Buballa\,\orcidlink{0000-0003-3747-6865}}
\email{michael.buballa@tu-darmstadt.de}
\affiliation{Technische Universit{\"a}t Darmstadt, Fachbereich Physik, Institut f{\"u}r Kernphysik,
Theoriezentrum, Schlossgartenstr. 2, D-64289 Darmstadt, Germany}
\affiliation{Helmholtz Forschungsakademie Hessen f{\"u}r FAIR (HFHF),
GSI Helmholtzzentrum f{\"u}r Schwerionenforschung,
Campus Darmstadt, D-64289 Darmstadt, Germany}

\author{H.~Gholami\,\orcidlink{0009-0003-3194-926X}}
\email{mohammadhossein.gholami@tu-darmstadt.de}
\affiliation{Technische Universit{\"a}t Darmstadt, Fachbereich Physik, Institut f{\"u}r Kernphysik,
Theoriezentrum, Schlossgartenstr. 2, D-64289 Darmstadt, Germany}
 
 \author{A.~Haber\,\orcidlink{0000-0002-5511-9565}}
 \email{a.haber@soton.ac.uk}
 \affiliation{Mathematical Sciences and STAG Research Centre,
University of Southampton, Southampton SO17 1BJ, United Kingdom}
 \affiliation{Department of Physics, Washington University in St.~Louis, St.~Louis, MO 63130, USA}

\author{M.~Hofmann\,\orcidlink{0000-0002-4947-1693}}
\email{marco.hofmann.physics@protonmail.com (Corresponding Author)}
\affiliation{Technische Universit{\"a}t Darmstadt, Fachbereich Physik, Institut f{\"u}r Kernphysik,
Theoriezentrum, Schlossgartenstr. 2, D-64289 Darmstadt, Germany}
\affiliation{Department of Physics, Washington University in St.~Louis, St.~Louis, MO 63130, USA}

\begin{abstract}
We calculate the absorption mean free paths of electron and muon neutrinos 
in two-flavor color-superconducting (2SC) quark matter in the density and temperature range that is relevant to binary neutron star mergers.
We model the strong interaction between quarks using a Nambu--Jona-Lasinio model, performing calculations self-consistently in the mean-field approximation. Since the 2SC gap is large we restrict our analysis to the contribution of unpaired quarks. We find that at low temperatures absorption by a down quark $\nu+d \to u+e^-/\mu^-$ is kinematically not allowed, so absorption by a strange quark $\nu+s \to u+e^-/\mu^-$ dominates the mean free path. As temperature or neutrino energy rises, the $d$ quark absorption channel becomes active, and the mean free path shrinks. We find that in equilibrated 2SC matter with an electron lepton fraction $Y_{L_e}=0.1$, the neutrinos form a degenerate gas with a mean free path of meters or less, independent of the temperature.
\end{abstract}

\maketitle

\section{Introduction}
\label{sec:intro}

Neutron star mergers provide a unique laboratory for studying the dynamics of kilometer-sized volumes of matter interacting via all four fundamental forces at densities above
nuclear saturation density ($n_{\text{sat}}=0.16\,\fm^{-3}$) and temperatures up to about $100\,\MeV$~\cite{Perego:2019adq}.
 Neutrinos are an important contributor to this physics:
as the most weakly interacting particles they have the longest mean free path, and therefore play an important role in transport.
Their interactions with the medium strongly depend on the local thermodynamic conditions, such that their mean free path can cover the range from being much smaller than the merger region~\cite{Espino:2023dei,Pajkos:2024iry,Ng:2024zve}, in which case they behave diffusively (``trapped'', contributing to thermal conductivity,
shear viscosity, etc.), to being much larger than the merger region, in which case they behave radiatively (``free streaming'', carrying off energy and lepton number) \cite{Most:2022yhe, Alford:2023gxq}. See also Ref.~\cite{Alford:2026kwd} for an analysis of neutrino trapping and free streaming using data from binary neutron-star simulations with neutrino transport.
See Refs.~\cite{Schmitt:2017efp,Foucart:2022bth} for review articles on neutrino transport in neutron stars and neutron star mergers.

In this paper, we make a comprehensive study of the absorption mean free path of neutrinos in one of the exotic high-density phases that have been proposed
as a possible constituent of neutron star cores, the two-flavor color-superconducting (2SC) phase. This phase is projected to arise at moderate densities where the density of strange ($s$) quarks is suppressed by their mass~\cite{Alford:1999pa}. 
The 2SC phase features Cooper pairing between the up ($u$) and down ($d$) quarks of two of the colors, which by convention we will identify as ``red'' and ``green''.  The pairing leads to an energy gap estimated to be in the $100-300\,\MeV$ range for those species~\cite{Braun:2021uua,Braun:2022jme,Geissel:2024nmx}.
We will calculate the density and temperature dependence of the quark masses, diquark gaps and thus the phase structure, and the neutrino mean free path self-consistently using a Nambu--Jona-Lasinio (NJL) model for the strong interaction. Within this model, we address the following question: at which densities and temperatures should neutrinos, depending on their energy, be treated as free streaming, and at which densities and temperatures is it relevant to consider effects of neutrino trapping? We study densities up to about $4\,\nsat$ and temperatures up to $50\,\MeV$, which are within the range of conditions found in neutron star mergers where our model predicts the 2SC phase would be the favored phase of quark matter.
The transition density from hadronic matter to quark matter depends on the hadronic equation of state and the bag constant. Given that our results are independent of these unknown parameters, we do not make any choices and present results valid for the quark matter phase, however far down in density it might reach.

Neutrino scattering and absorption have previously been studied in unpaired quark matter.
A commonly used approach is the bag model, where the quarks may be treated as free  \cite{Steiner:2001rp}
or have strong interactions calculated perturbatively
\cite{Iwamoto:1982zz,Gupta:1995bx,Colvero:2014zfa}.
If the quarks are free, then in the low temperature limit, there is no
kinematically allowed
phase space for neutrino absorption.
If perturbative strong interactions are included 
\cite{Iwamoto:1982zz,Gupta:1995bx,Wang:2006tg,Colvero:2014zfa} 
the resultant  corrections \cite{Baym:1975va,Pal:2011ve,Adhya:2012sq} create some
available phase space. This illustrates that the calculated rates will be dependent on how strong interactions are treated. In the present work, using an NJL model, strong interactions affect the rate via the mean-field corrections to the
particle masses and energy shifts.

Weak interaction processes in color-superconducting quark matter have been studied mostly in the context of neutrino emission and neutron star cooling \cite{Blaschke:1999qx,Jaikumar:2001hq,Jaikumar:2002vg,Alford:2004zr,Grigorian:2004jq,Jaikumar:2005hy,Schmitt:2005wg,Anglani:2006br,Negreiros:2012, Berdermann:2016mwt}, bulk viscosity \cite{Alford:2006gy,Wang:2010ydb,Berdermann:2016mwt,Alford:2024tyj,Alford:2025tbp} and neutrino mean free paths \cite{Carter:2000xf,Reddy:2002xc,Kundu:2004mz}. Ref.~\cite{Carter:2000xf} focused on the temperature dependence of pair-breaking effects
assuming massless quarks, while Refs.~\cite{Reddy:2002xc,Kundu:2004mz} included neutrino processes involving Goldstone bosons in the color-flavor locked (CFL) phase.

In the present paper, the density and temperature dependent phase space is determined self-consistently within the NJL model. In this model, unlike the perturbatively corrected bag model, the phase space for neutrino absorption by a down quark remains kinematically blocked at low temperatures by a small momentum deficit. As the temperature rises above this deficit, the channel is opened thermally (see Sec.~\ref{sec:phase_space}).

We calculate the mean free path of neutrinos due to absorption by the unpaired (blue) down or strange quarks. Specifically, the following processes: 
\begin{align}
    \nu_l+d_b&\to l^- + u_b \label{eq:ud_process}\ , \\
    \nu_l+s_b&\to l^- + u_b \label{eq:us_process}\ ,
\end{align}
which we will refer to as ``$d$-quark capture'' and ``$s$-quark capture'' respectively.
In the first (second) process, the neutrino of lepton flavor $l=e,\mu$ (for electron and muon, respectively) is absorbed by a blue down (strange) quark, emitting the lepton and a blue up quark.

We neglect processes involving the paired quarks: this includes neutrino absorption by a down quark that is red or green, but also absorption by the red or green strange quarks, since that would produce a red or green up quark. Such pair-breaking effects will be negligible at temperatures $T\ll T_c$, where they
are suppressed by  a factor $\sim\exp(-T_c/T)$ \cite{Jaikumar:2005hy} and $T_c$ is the critical temperature at which the 2SC condensate melts. In our model $T_c \gtrsim 100\,\MeV$, see Fig.~\ref{fig:pd_no_contours}, so pair-breaking will only become relevant
at temperatures above several tens of MeV,
where it would provide additional channels
leading to a shorter mean free path.
Since our calculations already predict that at such temperatures the mean free paths will be much smaller than a typical merger simulation fluid element, our conclusions about neutrino trapping would not be affected. We also neglect sub-dominant processes involving anti-leptons in the incoming channel, as they have a small phase space at low temperatures before the neutrinos get trapped via the processes studied in this paper. These include $d + l^+ + \nu_l \rightarrow u$ and $s + l^+ + \nu_l \rightarrow u$. While we expect similar results for anti-neutrino absorption opacities originating from processes like $u + \bar{\nu}_l \rightarrow d + l^+$ and $u + \bar{\nu}_l \rightarrow s + l^+$, we focus on the physics of neutrinos in this work. Three-to-one processes like $u + l^- + \bar{\nu}_l \rightarrow d$ and $u + l^- + \bar{\nu}_l \rightarrow s$ are often slower compared to the two-to-two counterparts, but would also contribute to the anti-neutrino absorption opacities.

This paper is organized as follows: In Sec.~\ref{sec:methods} we introduce the NJL model, explain how the neutrino mean free paths are calculated, and give an estimate of the phase space for the absorption processes at zero temperature. 
After showing the phase diagram of the model in Sec.~\ref{sec:phase_diagram}, we present results for the absorption mean free paths of the neutrinos via $d$-quark capture  \eqref{eq:ud_process} and $s$ quark capture \eqref{eq:us_process}, for different densities and temperatures. In Secs.~\ref{sec:density_dependence} to \ref{sec:T_dependence} we focus on
neutrinos of energy $E_\nu=3T$, which would be the approximate average energy if the neutrinos
were thermally equilibrated with their surroundings but not trapped to form a degenerate gas.
We then relax this assumption and explore different neutrino energies, including a degenerate electron neutrino distribution in Sec.~\ref{sec:neutrino-spectrum}. Finally, we draw our conclusions in Sec.~\ref{sec:conclusion}.

We use natural units where $\hbar=c=k_B=1$.

\section{Methods}\label{sec:methods}

\subsection{Model for color-superconducting quark matter}

We model strong interactions between quarks, allowing for the possibility of Cooper pairing, via a Nambu--Jona-Lasinio model with the Lagrangian \cite{Rehberg:1995kh,Gastineau:2001zke,Klahn:2006iw}
\begin{align}
\label{eq:Lagrangian}
\mathcal{L}&=\bar{\psi}(i\slashed{\partial}+\gamma^0\hat{\mu}-\hat{m})\psi \nonumber\\
&+G_S\sum_{A=0}^8\left[(\bar{\psi}\tau_A\psi)^2 + (\bar{\psi} i \gamma_5 \tau_A \psi)^2\right] 
\nonumber\\
&- K \bigl[\text{det}_{\text{f}}\bigl(\bar{\psi}(\mathds{1}+\gamma_5)\psi\bigr) + \text{det}_{\text{f}}\bigl(\bar{\psi}(\mathds{1}-\gamma_5)\psi\bigr)\bigr] \nonumber\\[1ex]
&+ G_D
\big[\,\big(\bar{\psi}^a_\alpha i\gamma_5 \epsilon^{\alpha\beta\gamma} \epsilon_{abc}(\psi_C)^b_\beta\big)
\big((\bar{\psi}_C)_r^\rho i\gamma_5 \epsilon_{\rho\sigma\gamma} \epsilon^{rsc} \psi_s^\sigma\big)\nonumber\\
&\hspace{3em}+\; \big(\bar{\psi}^a_\alpha  \epsilon^{\alpha\beta\gamma} \epsilon_{abc}(\psi_C)^b_\beta\big)
\big((\bar{\psi}_C)_r^\rho  \epsilon_{\rho\sigma\gamma} \epsilon^{rsc} \psi_s^\sigma\big) \,\big]\nonumber \\
&-G_V\bigl(\bar\psi\gamma^{\mu}\psi\bigr)^2+\mathcal{L}_{\text{lep}}\ .
\end{align}
In this expression $\psi$ is the quark field with suppressed flavor $\alpha=u,d,s$ and color $a=r,g,b$ indices, $\hat{m}$ is the quark mass matrix, $\hat{\mu}$ is a matrix of chemical potentials, both diagonal in flavor space, $\tau_A$ are the Gell-Mann matrices in flavor space for $A=1,..,8$ 
complemented by $\tau_0=\sqrt{2/3}\,\mathds{1}_f$, $\psi_C\equiv C\bar{\psi}^T$ and $\bar{\psi}_C\equiv \psi^TC$
are the charge-conjugated spinors with the charge conjugation operator $C=i\gamma^2\gamma^0$. The Levi-Civita tensors $\epsilon^{\alpha\beta\gamma}$ and $\epsilon_{abc}$ are anti-symmetric in flavor space and color space, respectively.

The Lagrangian \eqref{eq:Lagrangian} contains four interaction terms with corresponding couplings. They are all invariant under $SU(3)_L\times SU(3)_R$ chiral flavor and $SU(3)$ color symmetry. 
The term proportional to $G_S$ (mass dimension -2) is the standard scalar and pseudoscalar NJL interaction, which induces dynamical chiral symmetry breaking in the model. The term proportional to the coupling $K$ (mass dimension -5) is a $U(1)_A$-breaking six-quark interaction motivated by instantons \cite{Kobayashi:1970ji,tHooft:1976rip}. The term proportional to $G_D$ (mass dimension -2) allows for quark pairing in the Lorentz scalar and pseudoscalar, color and flavor antitriplet channel, leading to the formation of color-superconducting condensates. In this work, we use the mean-field approximation for the NJL model and only consider condensation in the Lorentz scalar channel. A repulsive vector interaction proportional to the coupling $G_V$ (mass dimension -2) is added so that compact stars with quark matter cores can reach the highest observed mass
$\sim 2M_{\odot}$ \cite{Klahn:2006iw}. Electrons and muons are added to the system as a free relativistic Fermi gas. The conserved quantities in our model are baryon number, electric charge, and color charge. When not explicitly stated otherwise, the corresponding chemical potentials are then fixed by cold, neutrinoless beta equilibrium. The Lagrangian of the leptons is denoted $\mathcal{L}_{\text{lep}}$ in Eq.~\eqref{eq:Lagrangian}. 

The NJL model is non-renormalizable and is most conveniently regularized using a sharp three-momentum cutoff, which we fix as $\Lambda'=602.3\,\MeV$.
The values of the couplings, $G_S\,\Lambda'^2=1.835$, $K\,\Lambda'^5=12.36$ and the bare quark masses $m_{u,d}=5.5\,\MeV$ and $m_s=140.7\,\MeV$ are fitted to the meson spectrum
in vacuum \cite{Rehberg:1995kh}.
To avoid cutoff artifacts when the model is evaluated in the medium 
we use the renormalization-group consistent treatment of Ref.~\cite{Gholami:2024diy}, see also Ref.~\cite{Braun:2018svj}. 
Specifically, among the schemes discussed in Ref.~\cite{Gholami:2024diy}, we employ the renormalization-group consistent minimal 
scheme, motivated by the analogy to wave function renormalization of the diquark field~\cite{Gholami:2024diy,Gholami:2025afm}.

We fix the diquark coupling $G_D=1.49\,G_S$ and the vector coupling $G_V=0.77\,G_S$.
These values allow for a hybrid star with a hadron-quark phase transition fulfilling all astrophysical constraints \cite{Gholami:2024ety,Christian:2025dhe}. The ground state and spectrum of quark excitations as a function of
baryon chemical potential $\mu_B$ and temperature $T$ is obtained from the mean-field approximation to the effective potential, $\Omega_\text{eff}$. The diquark, chiral, and vector
condensates are obtained by 
solving the gap equations
and taking the solution with the lowest free energy.
Color and electric neutrality are imposed by extremizing $\Omega_\text{eff}$ with respect to
the corresponding chemical potentials. Details on this procedure and the renormalization-group consistent formulation are given in Refs.~\cite{Gholami:2024ety,Gholami:2024diy}. 
\subsection{Calculation of neutrino absorption mean free paths}
We use the labels $u$, $l$, and $\nu$ for the up quark, the lepton ($l=e,\mu$), and the electron or muon neutrino, respectively. The particle that absorbs the neutrino is either a down quark for $d$-quark capture \eqref{eq:ud_process} or a strange quark for $s$-quark capture \eqref{eq:us_process}. When discussing both processes in the same equation, we use the label $d/s$. 
The inverse mean free path for the absorption processes
\eqref{eq:ud_process} and \eqref{eq:us_process} are given by
\begin{align}
\label{eq:golden_rule}
\lambda^{-1}&=N_c\cdot G_F^2\int\frac{d^3\bm{p}_{d/s}}{(2\pi)^3}\int\frac{d^3\bm{p}_l}{(2\pi)^3}\int\frac{d^3\bm{p}_u}{(2\pi)^3}\\&\times(2\pi)^4\delta^4(P_\nu{+}P_{d/s}{-}P_l{-}P_u)
\mathcal{M} f_{d/s}(1-f_l)(1-f_u). \nonumber
\end{align}
In the 2SC phase, only the strange quarks and the blue up and down quarks are unpaired, so the only processes containing only unpaired quarks are \eqref{eq:ud_process} and \eqref{eq:us_process}; thus, in our calculation, we set the
number of colors \mbox{$N_c=1$}.
We denote the three-momentum of particle $i$ as $\bm{p}_i$ and four-momentum as $P_i$. The integral is over the three-momenta of the incoming down or strange quark and of the outgoing up quark and lepton. The delta-distribution imposes four-momentum conservation and $G_F=1.16637\cdot 10^{-11}\,\MeV^{-2}$ is the Fermi constant. The Fermi--Dirac distributions $f_i$ depend on the energy of the individual particles $E_i$, their chemical potentials $\mu_i$, and the temperature $T$. The factors $(1-f_l)$ and $(1-f_u)$ account for Pauli blocking of the final states. The dimensionless squared matrix element is summed 
over initial and final spins 
and is given by Ref.~\cite{Steiner:2001rp} 
\begin{align}
    \mathcal{M} =& \frac{2}{{E_\nu E_{d/s}^\ast E_l E_u^\ast}}\big[(\mathcal{V}+\mathcal{A})^2(P_\nu \cdot P_{d/s})( P_l \cdot P_u) \nonumber\\
    &+(\mathcal{V}-\mathcal{A})^2(P_\nu \cdot P_u)( P_l \cdot P_{d/s}) \nonumber\\
    &- (\mathcal{V}^2-\mathcal{A}^2)(P_\nu \cdot P_l)(M_{d/s} M_u)\big]\ ,
    \label{eq:matrixelement}
\end{align}
with $\mathcal{V}=\mathcal{A}=\cos(\theta_c)$ for $d$-quark capture \eqref{eq:ud_process} and $\mathcal{V}=\mathcal{A}=\sin(\theta_c)$ for $s$-quark capture \eqref{eq:us_process}, respectively, such that, unlike the situation in hadronic matter, only the first term in Eq.~\eqref{eq:matrixelement} is nonzero. Here, $E_i^\ast=\sqrt{p_i^2+M_i^2}$ denote the quark energies without any energy shifts due to the interaction and the angle $\theta_c=13.02^\degree$ is the Cabibbo angle.

Following Ref.~\cite{Gupta:1995bx} (see also Ref.~\cite{Steiner:2001rp}), the expression \eqref{eq:golden_rule} can be simplified to
\begin{align}
	\lambda^{-1}=&(\mathcal{V}+\mathcal{A})^2\frac{N_c G_F^2}{4\pi^3}\frac{1}{E_\nu p_\nu}\int_{E^\ast_{d/s,\text{min}}}^{\infty} dE_{d/s}^\ast \int_{M_l}^{E_{l,\text{max}}} dE_l \nonumber\\
    &\times f_{d/s}(1-f_l)(1-f_u)
	\Big[p^{\nu,d/s }p^{l,u} (Q_{\text{max}}-Q_{\text{min}}) \nonumber\\
    &-\frac{1}{6}(p^{\nu, d/s}+ p^{l,u})(Q_{\text{max}}-Q_{\text{min}})^3 \nonumber\\
    &+\frac{1}{20}(Q_{\text{max}}-Q_{\text{min}})^5 \Big]\Theta(Q_{\text{max}}-Q_{\text{min}})  \ ,\label{eq:opacity}
\end{align}
with $p^{i,j}=E_i E_j^\ast + \frac{1}{2}(p_i^2 + p_j^2)$,
\begin{align}
Q_{\text{min}}=&\text{max}\{\vert p_\nu-p_{d/s}\vert,\vert p_l-p_u \vert\} \, \\
Q_{\text{max}}=&\text{min}\{p_\nu+p_{d/s},p_l+p_u\} \,
\end{align}
with the Heaviside function $\Theta$ and the integration bounds $E^\ast_{d/s,\text{min}}=\text{max}\{M_{d/s},M_l+M_u-E_\nu\}$, and $E_{l,\text{max}}=E_\nu+E^\ast_{d/s}-M_u$ following from energy conservation. 
For numerical evaluation, we found that an upper bound on the $E^\ast_{d/s}$ integral of $E^\ast_{d/s,\text{max}}=\mu_{d/s}+E_\nu+15T$ was sufficient.

To evaluate the Fermi-Dirac distributions in the integral for the mean free path \eqref{eq:opacity}, we need the dispersion relations of the unpaired (blue) quarks of flavor $f\in\{u,d,s\}$, 
\begin{equation}
    E_{f}=E_f^\ast+6G_V n_B \ ,
\end{equation}
where the density and temperature dependent constituent masses $M_f$ are calculated from the gap equations and are shown for zero temperature in Fig.~\ref{fig:masses}.
Note that the vector interaction shifts the
energies by a term proportional to the net baryon density $n_B$  (see, e.g., Ref.~\cite{Buballa:2003qv}).

The chemical potentials of the blue quarks are
\begin{align}
     \mu_{ub} &= \frac{\mu_B}{3} + \frac{2}{3}\mu_Q -\frac{2}{\sqrt{3}}\mu_8 \, ,\\
     \mu_{db}= \mu_{sb} &= \frac{\mu_B}{3} - \frac{1}{3}\mu_Q -\frac{2}{\sqrt{3}}\mu_8 \ .
\end{align}
The electric charge chemical potential $\mu_Q$ and the color charge chemical potentials $\mu_3$ and $\mu_8$ are determined by electric charge neutrality and color neutrality, respectively. We assume that the non-leptonic flavor-changing rates are fast compared to the relevant timescales such that $\mu_{db}=\mu_{sb}$. For convenience, we define the effective chemical potentials for quarks of flavor $f$ and color $c$,
\begin{equation}\label{eq:effmudef}
    \tilde{\mu}_{fc}=\mu_{fc}-6G_Vn_B \ .
\end{equation}

\subsection{Fermi momenta and phase space}\label{sec:phase_space} 

The mean free path is strongly influenced by the available phase space for the $d$-quark capture \eqref{eq:ud_process} and $s$-quark capture \eqref{eq:us_process} processes.
At temperatures and neutrino energies much smaller than the Fermi energies of the quarks and leptons, only particles close to their Fermi surface can participate in the scattering process. In this limit, the phase space is constrained by the following triangle inequalities,
\begin{align}
   p_F^u + p_F^l - p_F^{d/s} &\geq 0 \label{eq:ineq_1} \ ,\\
   p_F^{d/s} + p_F^u - p_F^{l} &\geq 0 \label{eq:ineq_2}\ ,\\
   p_F^l + p_F^{d/s} - p_F^{u} &\geq 0 \label{eq:ineq_3}\ ,
\end{align}
where the Fermi momentum of an unpaired (blue) quark with flavor $f$ is
\begin{equation}
    p_F^{f}=\sqrt{\tilde{\mu}_{fb}^2-M_f^2} \ .
\end{equation}

If any of the triangle inequalities cannot be fulfilled, then the phase space available for the process is strongly suppressed in the limit $T,E_\nu\to 0$,
leading to a long mean free path.
As we will see, the triangle inequalities play an essential role in determining the dependence of neutrino mean free paths on density and temperature. We now discuss how they influence the rates of the two relevant processes, $s$-quark capture and $d$-quark capture.

The $s$ quark absorption process becomes relevant when the strange quark effective chemical potential rises above the strange quark mass; in our model, this happens at densities $n_B \gtrsim 3.2\,\nsat$, see Sec.~\ref{sec:results}.
At those densities, all three triangle inequalities \eqref{eq:ineq_1}, \eqref{eq:ineq_2}, and \eqref{eq:ineq_3} are obeyed.

Unlike the strange quarks, down quarks are available at all densities. In $d$-quark capture, the two triangle equalities \eqref{eq:ineq_2} and \eqref{eq:ineq_3} are always obeyed, but
\eqref{eq:ineq_1} is never obeyed. Therefore, we refer to $p_F^d - p_F^{u} - p_F^l>0$ as the momentum deficit, which becomes smaller as the density rises. To illustrate this,
we assume $M_u\approx M_d \ll \tilde{\mu}_{u,d}$ 
(which we will see in Sec.~\ref{sec:results} is generally a good approximation) and omit the color index, because only blue quarks are considered. 
The momentum deficit becomes
\begin{align}
    p_F^d -p_F^u-p_F^e 
	&\approx \mu_d - \mu_u - \mu_e + \frac{1}{2}\left(\frac{M_u^2}{\mu_u}-\frac{M_d^2}{\mu_d}\right)\label{eq:momentum_deficit} \\
    &\approx -\mu_Q\frac{M^2}{2\mu_u \mu_d} \ , \label{eq:momentum_deficit_approx}
\end{align}
where $M$ is the common mass of $u$ and $d$ quarks.
Here, we assumed beta equilibrium with no trapped neutrinos, $\mu_u+\mu_e=\mu_d$.
In neutral matter, $\mu_Q=\mu_u-\mu_d<0$ so the momentum deficit is always positive, meaning that the process is suppressed, but only by a small amount which gets smaller as
the density rises\footnote{We note that $\mu_Q$ is more negative in 2SC matter compared to unpaired quark matter, such that the $d$-quark and $s$-quark capture phase space for unpaired quarks in the 2SC phase is typically smaller than for unpaired quarks in normal quark matter.}. As we will see in Sec.~\ref{sec:results}, when this deficit becomes small enough, thermal blurring of the Fermi surfaces, or momentum injected by the incoming neutrino, can be sufficient to relieve some of the suppression of the rate.

Note that \eqref{eq:momentum_deficit} and \eqref{eq:momentum_deficit_approx} do not contain the quark energy shift due to the vector interaction, because in our model these are the same for every flavor and hence cancel out. In a model with a flavor-dependent energy shift, this cancellation would not occur. We discuss this briefly in Sec.~\ref{sec:conclusion}, and the analogous phenomenon for perturbatively-corrected bag models in Appendix~\ref{sec:app}, but leave a study with a flavor-dependent vector interaction for future work.

\section{Results}
\label{sec:results}

In Sec.~\ref{sec:phase_diagram}, we show what the NJL model predicts for the phase diagram and the chemical composition and mean fields calculated at zero temperature. The next three subsections contain 
results for the mean free path of neutrinos
of energy $E_\nu=3T$, the average energy in a nondegenerate thermal population.  In Sec.~\ref{sec:density_dependence} and \ref{sec:thermal-phasediagram}, we show the density dependence of the absorption mean free paths of electron and muon neutrinos at different temperatures, distinguishing the
contributions of
the $d$-quark capture and $s$-quark capture processes. The temperature dependence is studied in Sec.~\ref{sec:T_dependence}. Then in Sec.~\ref{sec:neutrino-spectrum} we generalize to
arbitrary neutrino energy, focusing on 2SC matter with
electron lepton fraction $Y_{L_e}=0.1$.

\subsection{2SC matter in the phase diagram}\label{sec:phase_diagram}

\begin{figure}
    \centering
\includegraphics[width=\linewidth]{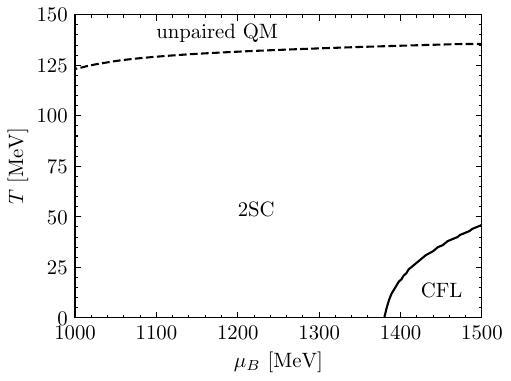}\\
\includegraphics[width=\linewidth]{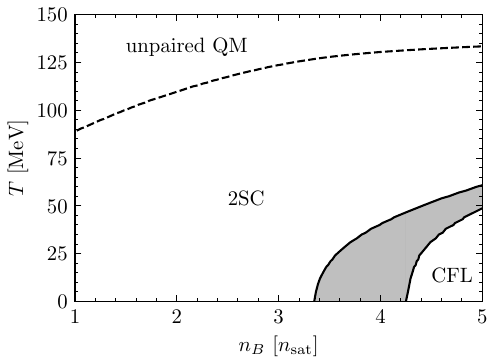}
\caption{
Phase diagram for locally neutral matter in cold beta equilibrium described by the NJL model \eqref{eq:Lagrangian}. Top panel: the baryon chemical potential $\mu_B$ and temperature $T$ plane. Bottom panel: the baryon density and temperature plane.
The 2SC-CFL phase transition is first order (solid lines), and the transition of 2SC matter to unpaired quark matter is second order (dashed lines). The gray-shaded area represents the jump in number density at the first-order phase transition between the two phases.
}
  \label{fig:pd_no_contours}
\end{figure}

The phase diagram of locally neutral matter in cold beta equilibrium described by the NJL model is shown in Fig.~\ref{fig:pd_no_contours}, in the plane spanned by baryon chemical potential $\mu_B$ and temperature $T$ (upper panel) and in the plane spanned by the net baryon density $n_B$ and temperature (lower panel). Since 2SC matter is favored over the CFL phase at densities up to about $3.4\,\nsat$, our calculations cover that range. 
At higher chemical potentials, corresponding to $n_B \gtrsim 4.2\,\nsat$, we have a CFL phase, in which quarks of all colors and flavors are paired.
We choose $n_B=\nsat$ as the lower bound for our calculations, but it should be noted that the NJL model does not include confinement, so the 2SC phase might be replaced by a hadronic phase before that lower bound is reached.

\begin{figure}
    \centering
\includegraphics[width=\linewidth]{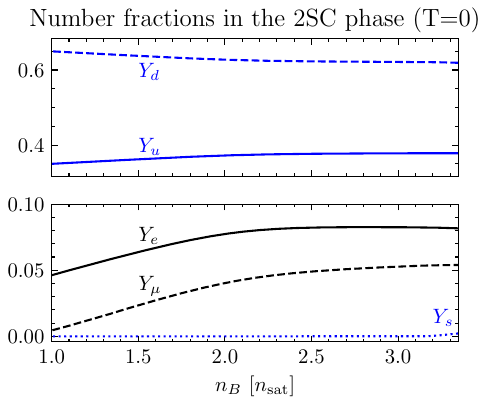}
    \caption{
    Number fractions $Y_u$,$Y_d$ and $Y_s$ of up, down and strange quarks, respectively, in the 2SC phase at zero temperature. The black lines in the bottom panel refer to the number fraction of electrons $Y_{e}$ and muons $Y_\mu$.
    }
\label{fig:composition}
\end{figure}

Fig.~\ref{fig:composition} shows the net particle number fractions $Y_i\equiv n_i/n_B$ for up, down, and strange quarks, and electrons and muons, as a function of density at zero temperature. The requirement of electrical neutrality
ensures that approximately 1/3 of the quarks are up quarks and the remaining 2/3 are down quarks. 
Strange quarks appear at $n_B\gtrsim 3.2\,\nsat$, but $Y_s$ never exceeds $1\%$ in the 2SC phase at zero temperature.
 
The electron fraction $Y_e$ has a maximum value around 0.08, which implies that if 
neutrinos were trapped, there would be a net positive electron neutrino population for electron lepton fraction $Y_{L_e}\equiv Y_e+Y_{\nu_e}$ above $0.08$, which is at the upper end of the range expected in the central region in mergers \cite{Perego:2019adq,Pajkos:2024iry}. At a lower electron lepton fraction there would be a significant antineutrino population, motivating a study of antineutrino mean free paths. We leave this for future work.

\begin{figure}
    \centering
\includegraphics[width=\linewidth]{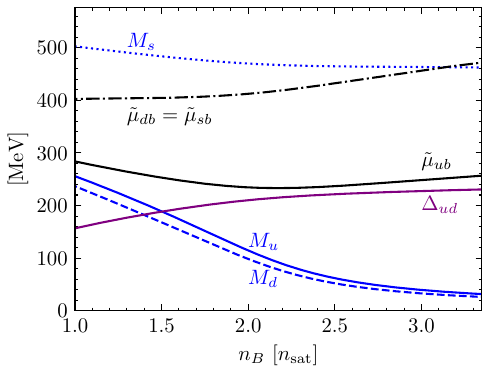}
    \caption{
    Constituent quark masses $M_u, M_d$, and $M_s$, effective chemical potentials $\tilde{\mu}_{ub},\tilde{\mu}_{db}$, and $\tilde{\mu}_{sb}$ of the up, down, and strange quarks, respectively, and diquark pairing gap $\Delta_{ud}$ plotted over the net baryon density in units of the nuclear saturation density at zero temperature. 
    }
    \label{fig:masses}
\end{figure}

In Fig.~\ref{fig:masses} we show the effective chemical potentials $\tilde\mu$ (Eq.~\eqref{eq:effmudef}) and constituent masses $M$ for the relevant quark species, along with the 2SC pairing gap $\Delta_{ud}$.
Because of electrical neutrality, the effective chemical potential for the down quarks is more than $100\,\MeV$ larger than the effective chemical potential for the up quarks in the 2SC phase. The up and down constituent quark masses are very similar and become much smaller than the effective chemical potentials with increasing density, justifying the small mass approximation used to obtain Eq.~\eqref{eq:momentum_deficit_approx}. The onset of the strange quarks is at $n_B\approx 3.2\,\nsat$, when $\tilde{\mu}_s>M_s$.

\begin{figure*}[t]
    \centering
    \includegraphics[width=\linewidth]{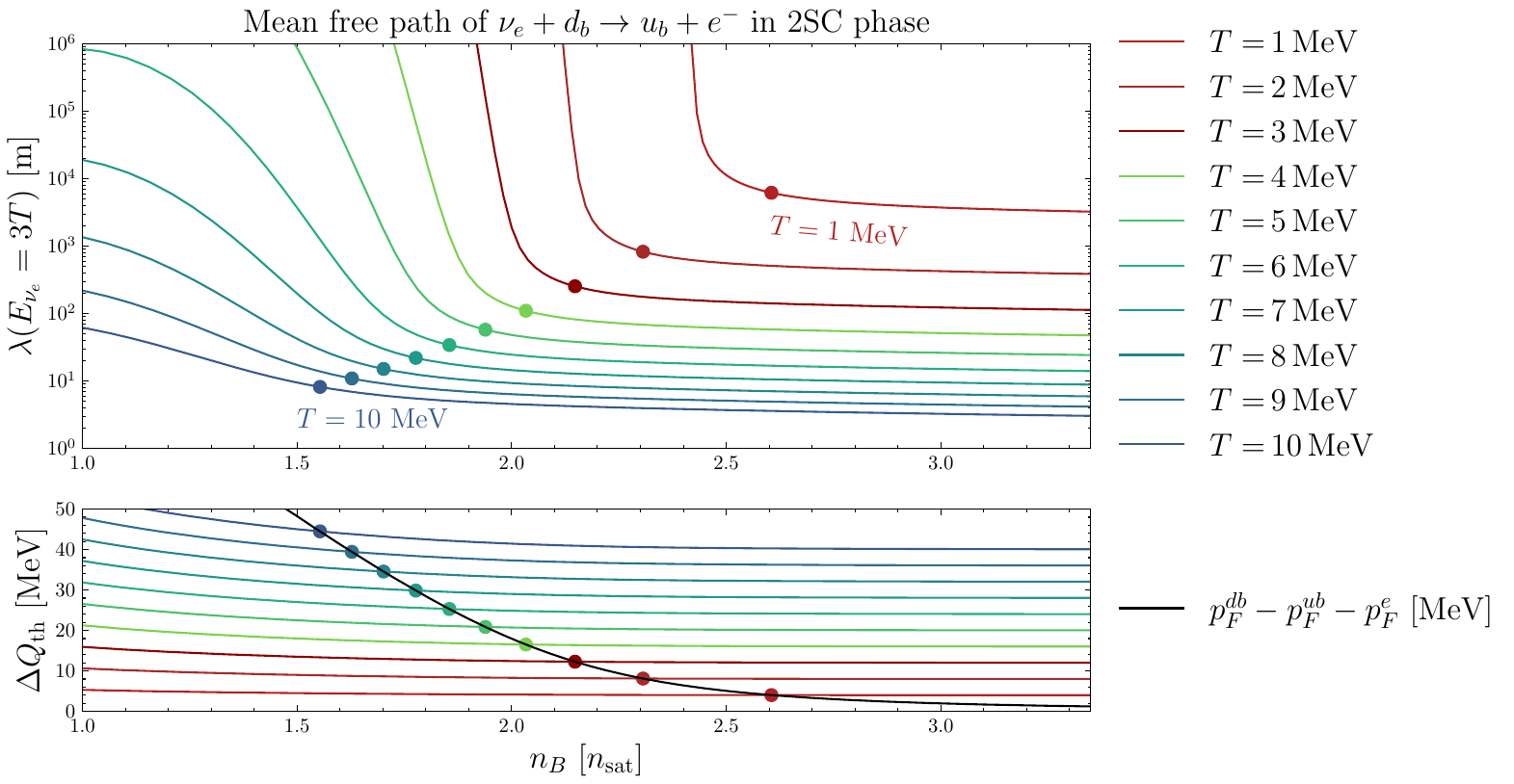}
 \caption{
 Top: electron neutrino mean free path for $d$-quark capture of a neutrino with energy $E_{\nu}= 3T$, as a function of baryon density $n_B$.
 Different colors correspond to different temperatures $T$. Bottom: momentum deficit $p_F^{db} - p_F^{ub}-p_F^e$ (black) and lines of additional thermal phase space $\Delta Q_{\text{th}}$. The densities where the momentum deficit and $\Delta Q_{\text{th}}$ intersect (dots in both panels) give an estimate of the densities beyond which the process is not suppressed.}
\label{fig:density_dependence_down}
\end{figure*}

\begin{figure*}[t]
    \centering
    \includegraphics[width=\linewidth]{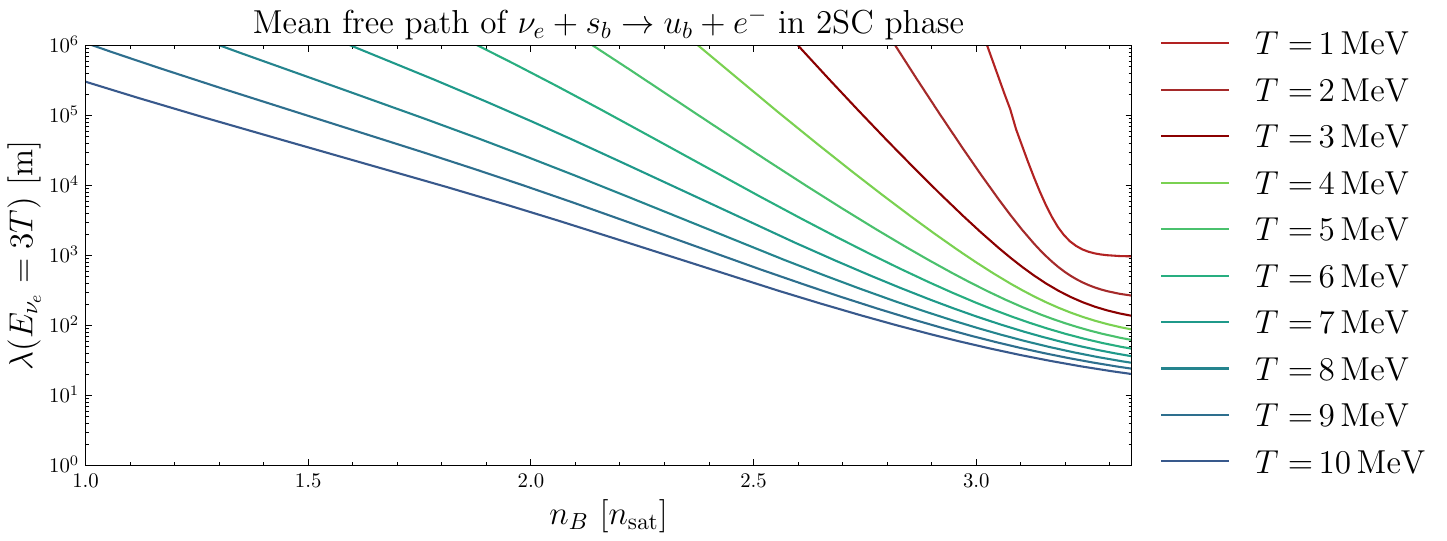}
    \caption{Electron neutrino mean free path for $s$-quark capture of a neutrino with energy $E_{\nu}= 3T$, as a function of baryon density $n_B$. The process is kinematically allowed, but the strange quark fraction is heavily suppressed at $n_B < 3.2\,\nsat$.}
\label{fig:density_dependence_strange}
\end{figure*}

\subsection{Density dependence of neutrino absorption}\label{sec:density_dependence}

We now survey our results for neutrino absorption mean free paths.  
We first focus on ``thermal'' neutrinos, i.e., those with energy $E_\nu\approx3T$. This would be the average neutrino energy if the neutrinos were non-degenerate but thermally equilibrated with the other degrees of freedom. We first show results for absorption of thermal electron neutrinos by down quarks, then by strange quarks, then the total.
We then show the corresponding results for muon neutrinos.

\subsubsection{Thermal neutrino absorption by down quarks}
\label{sec:thermal-down}

The density and temperature dependence of the neutrino absorption mean free path is shown in the upper panel of Fig.~\ref{fig:density_dependence_down} for the $d$-quark capture process with electrons $\nu_e+d_b\to u_b+e^-$. The different colored lines correspond to different temperatures.   For each temperature, we plot the mean free path of neutrinos with the typical thermal energy  $E_\nu=3T$.
As we discussed in Sec.~\ref{sec:phase_space}, the 
$d$-quark capture process is kinematically disallowed at zero temperature, due to a small momentum deficit (Eq.~\eqref{eq:momentum_deficit_approx}).
This results in long neutrino mean free paths at low temperatures and low densities.
However, as the temperature rises, more phase space starts to become available as the deficit is overcome by a combination of thermal blurring of the Fermi surfaces and the increasing momentum of the incoming neutrino. As a result, at each temperature there is an effective threshold density $n_B^\ast(T)$, at which the process \eqref{eq:ud_process} 
proceeds faster, producing a rapid decrease in the mean free path, after which it decreases more slowly as $\lambda \propto (p_F^e \cdot p_F^{db} \cdot p_F^{ub})^{-1} \propto n_B^{-1}$ at higher densities. This effective threshold is more pronounced at lower temperatures.

We can predict the threshold density directly from the equation of state at $T=0$  by comparing the  momentum deficit for $d$-quark capture,  $p_{F}^{db}-p_{F}^{ub}-p_{F}^e$,  with the  momentum $\Delta Q_{\text{th}}$ provided by thermal blurring and the incoming thermal neutrino.
In momentum space, thermal blurring allows  quarks to occupy states within a width of $\Delta p\sim T/v_F$ around their Fermi surfaces, where $v_F=(\partial E / \partial p)^{-1}\vert_{p=p_F}$ is the Fermi velocity. We therefore estimate
\begin{equation}
    \Delta Q_{\text{th}}(n_B,T)=\dfrac{T}{v_F^{ub}(n_B)} + \langle E_\nu\rangle(T),
    \label{eq:Qth}
\end{equation}
where $v_F^{ub}$ is the Fermi velocity of the blue up quark, which is close to 1 across the density range studied here, and $\langle E_\nu \rangle (T)=3T$ is the mean thermal neutrino energy.

In the bottom panel of Fig.~\ref{fig:density_dependence_down} we compare the exactly calculated zero-temperature momentum deficit $p_{F}^{db}-p_{F}^{ub}-p_F^e$ (solid black line)
with $\Delta Q_{\text{th}}$ (colored lines, one for each temperature).
$\Delta Q_{\text{th}}$ shows very mild density dependence, arising from its dependence on $v_F^{ub}$.
The colored dots show where $\Delta Q_{\text{th}}$ equals the momentum deficit, which
gives us a rough estimate of the density $n_B^{\ast}(T)$ at which the process is not suppressed any more. The dots are also reproduced on the mean free path curves in the upper panel, and we see that indeed they give a reasonable estimate of the density
at which the change to a shorter mean free path occurs.

\subsubsection{Neutrino absorption by strange quarks}
\label{sec:thermal-strange}
In Fig.~\ref{fig:density_dependence_strange}
we show the absorption mean free path
for the $s$-quark capture process $\nu_e+s_b\to u_b+e^-$.
This process is kinematically allowed at all densities; however, at densities below $3.2\,\nsat$, the strange quark chemical potential
is less than the constituent mass (Fig.~\ref{fig:composition}), so the
strange quark density is
exponentially suppressed as $\exp(-(M_s-\tilde\mu_s)/T)$, resulting in a long mean free path. As the density rises towards $3.2\,\nsat$, the suppression becomes less severe, and when $\tilde\mu_s>M_s$, there is a Fermi sphere of strange quarks, and the mean free path varies more slowly as $\lambda \propto n_B^{-1}$.

\subsubsection{Total neutrino absorption}
\label{sec:thermal-total}

\begin{figure*}[t]
    \centering
    \includegraphics[width=\linewidth]{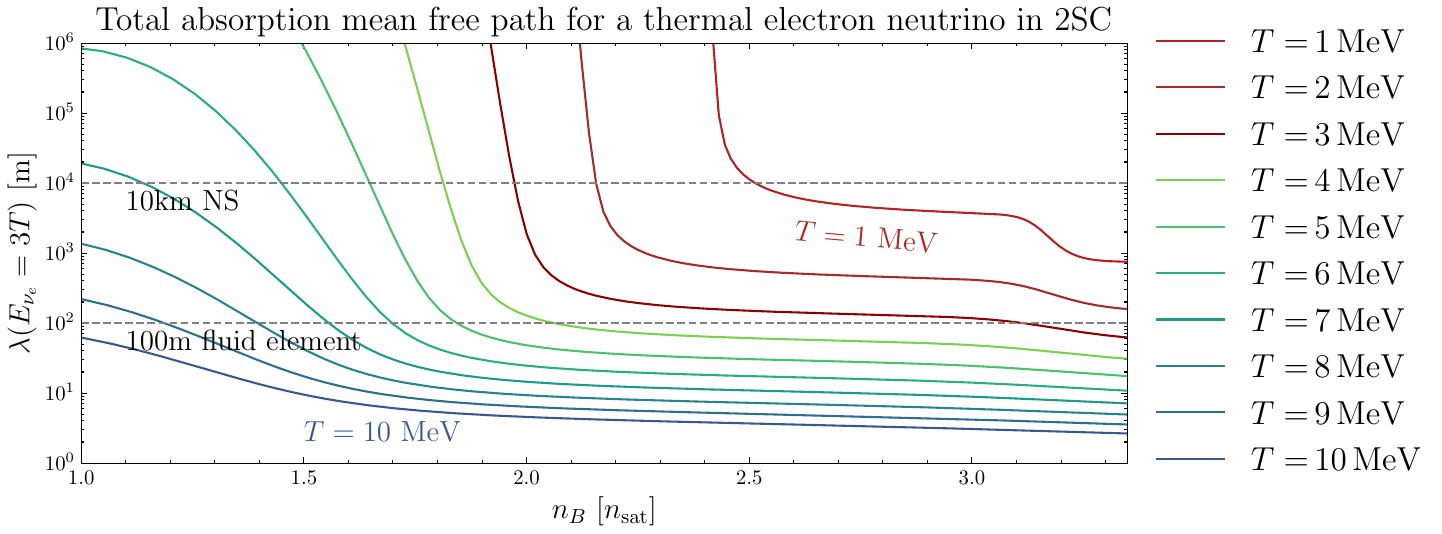}
    \caption{Total absorption mean free path of an electron neutrino of energy $E_\nu=3T$ in 2SC matter, as a function of density.
   Different colors correspond to different temperatures $T$.}
\label{fig:density_dependence}
\end{figure*}

\begin{figure*}[t]
    \centering
\includegraphics[width=\linewidth]{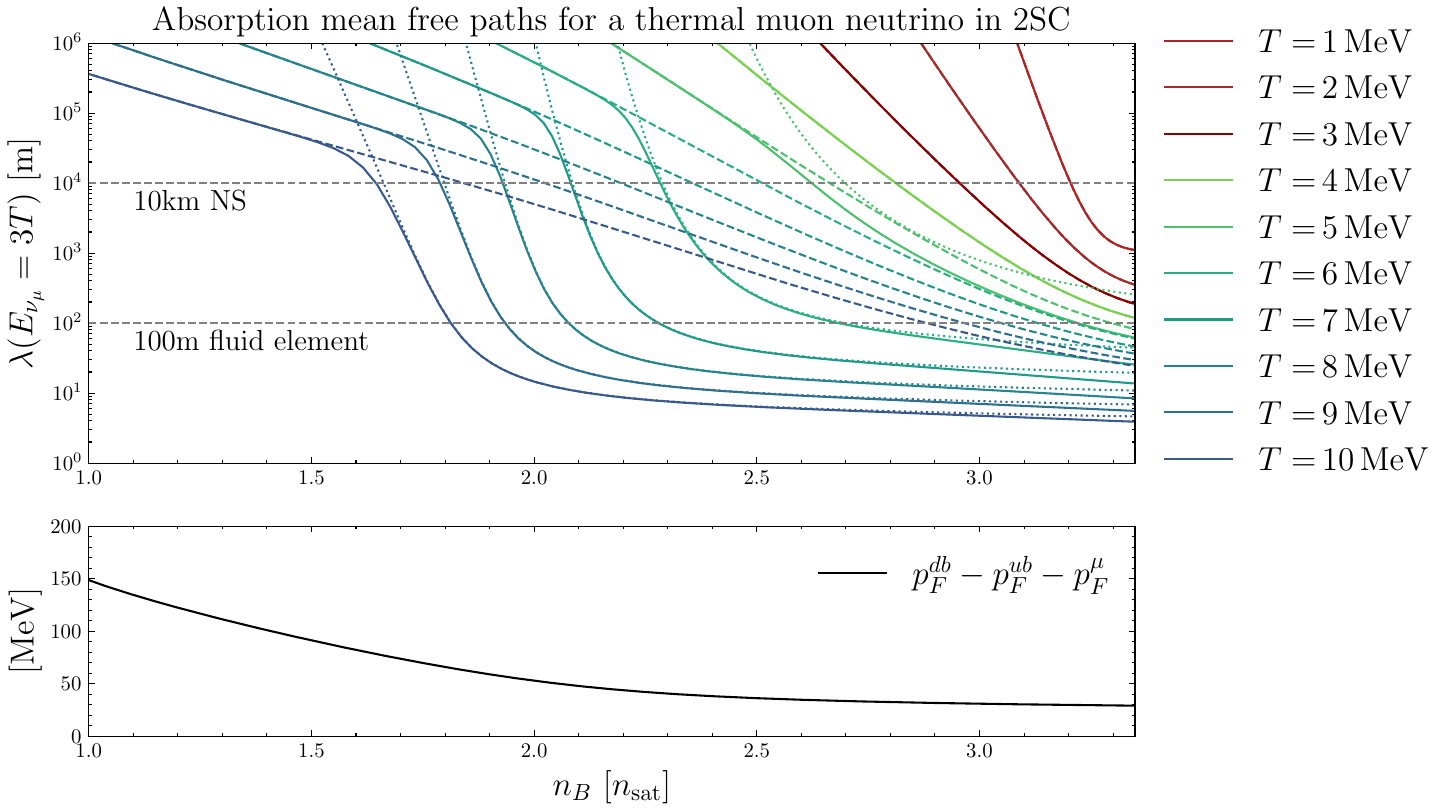}
    \caption{Top panel: muon neutrino absorption mean free path due to $d$-quark capture (dotted), $s$-quark capture  (dashed), and the total mean free path (solid), as a function of density.
    We assume a neutrino of average thermal energy $E_\nu=3T$.
    Different colors correspond to different temperatures $T$. Lower panel: momentum deficit for absorption by down quarks at $T=0$.}
\label{fig:muons}
\end{figure*}

The total absorption mean free path for neutrinos of energy $E_\nu=3T$ 
is obtained by adding the opacities, $\lambda_{\text{total}}=(\lambda_{d\to u}^{-1}+\lambda_{s\to u}^{-1})^{-1}$. The results are shown in Fig.~\ref{fig:density_dependence}. 
Dashed lines indicate distance scales that are relevant in neutron star mergers and simulations: 10\,km for the radius of a neutron star and 100\,m for the size of a typical fluid element in current simulations~\cite{Foucart:2024npn}.

We note the following features:
\begin{enumerate}
\item {\em Free streaming.}
The mean free path decreases monotonically with increasing baryon density and temperature. Thus at sufficiently low densities or temperatures the neutrinos can be treated as free-streaming on the astrophysical scales relevant to neutron stars.
\item {\em Trapping.}
Above some density, which decreases with increasing temperature due to thermal blurring, the mean free path of a thermal neutrino becomes shorter than these astrophysical distance scales, so neutrinos can no longer be considered as free-streaming. When the mean free path is sufficiently short the neutrinos become a trapped equilibrated gas, so conservation of lepton number becomes a relevant constraint.
We will discuss the mean free paths of such an equilibrated neutrino distribution in Sec.~\ref{sec:neutrino-spectrum}.
\item {\em Role of strange quarks}.
At densities below $3.2\,\nsat$, the mean free path is dominated by $d$-quark capture. However, as the density rises past $3.2\,\nsat$
we see a downward step in the mean free path  which becomes more noticeable at lower temperatures.
This arises from the onset of a degenerate population of strange quarks
at $n_B\approx3.2\,\nsat$, which allows $s$-quark capture to contribute to the total mean free path $\lambda_{\text{total}}$.  At low temperatures ($T\lesssim 3\,$MeV)
this processes dominates because it is always kinematically allowed, whereas $d$-quark capture is strongly suppressed at low temperatures.
\end{enumerate}

\subsubsection{Absorption of muon neutrinos}
Our results for the mean free path of muon neutrinos 
of energy $E_\nu=3T$ are plotted in
Fig.~\ref{fig:muons}, showing the total (solid line) and the separate contributions from absorption by down quarks (dotted), and strange quarks (dashed). The lower panel shows the momentum deficit for $d$-quark capture. This is
larger than for $d$-quark capture of electron neutrinos because the muon is heavier than the electron, giving it a smaller Fermi momentum.
Thus higher temperatures are needed to open phase space for the process. Indeed, for $T \lesssim 4\,\MeV$ and densities below the strange quark onset ($n_B\lesssim 3.2\,\nsat$),
the muon neutrinos are free streaming on the distance scales relevant to neutron stars. The $d$-quark capture process only
becomes dominant at $T \gtrsim 5\,\MeV$. As already seen in Fig.~\ref{fig:density_dependence_down} for the electron neutrinos, at these higher temperatures, there is again an effective temperature-dependent threshold density at which the mean free paths drop and the muon neutrinos leave the free-streaming regime.

\begin{figure}
    \centering
\includegraphics[width=\linewidth]{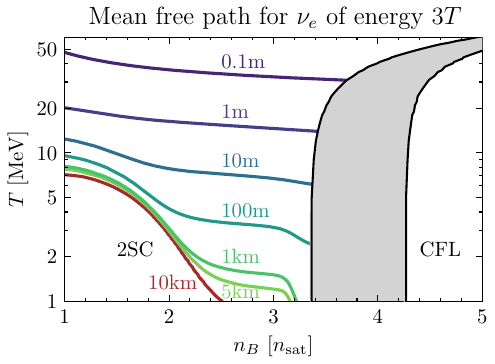}\\
\includegraphics[width=\linewidth]{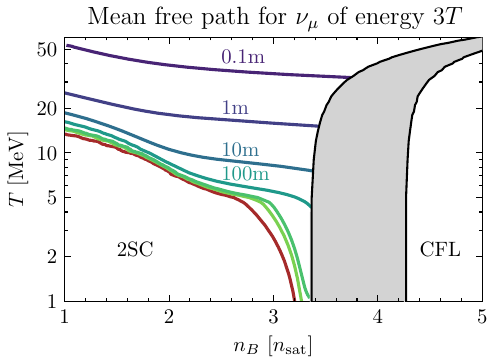}
\caption{Contour plot of the absorption mean free path in 2SC matter for electron neutrinos (top) and muon neutrinos (bottom) with the average thermal energy $E_\nu=3T$, displayed in the plane spanned by baryon density and temperature.}
\label{fig:pd}
\end{figure}

\begin{figure*}[htbp]
    \centering
\includegraphics[width=\linewidth]{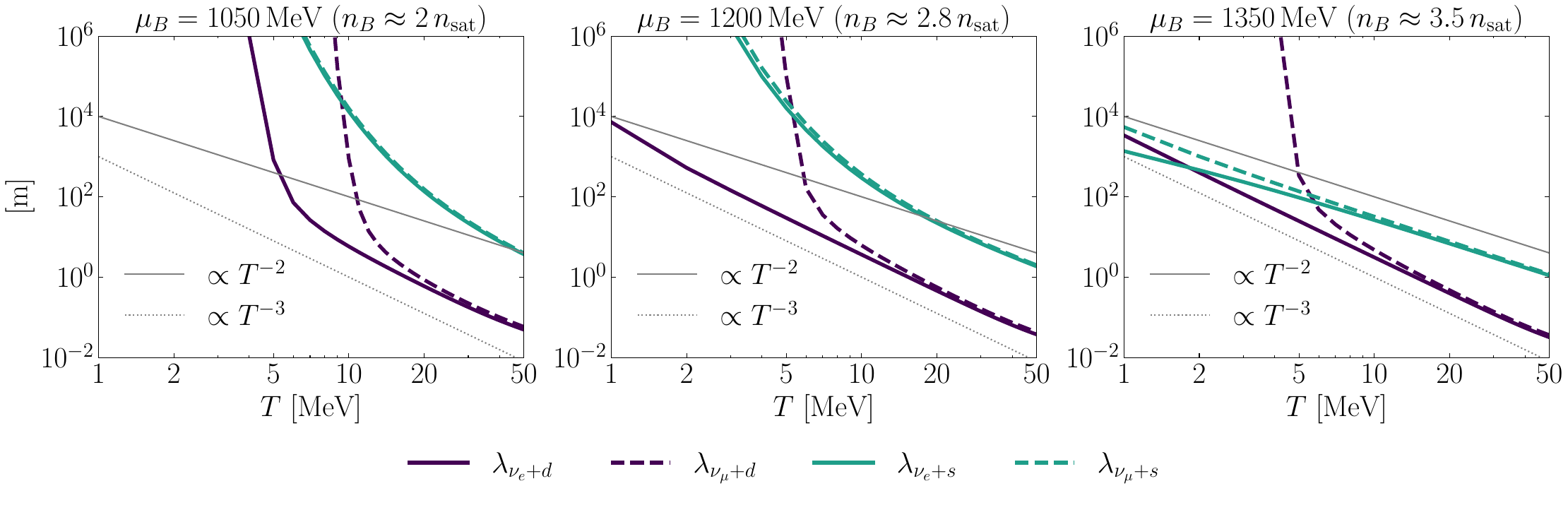}
    \caption{
    Temperature dependence of the neutrino absorption mean free paths of all considered processes at three different baryon chemical potentials. The neutrino energy is set to $E_\nu=3T$. For comparison, the thin solid and dotted lines illustrate $T^{-2}$ and $T^{-3}$ behavior, respectively.}
\label{fig:T_dependence}
\end{figure*}

\begin{figure*}[htbp]
    \centering
\includegraphics[width=\linewidth]{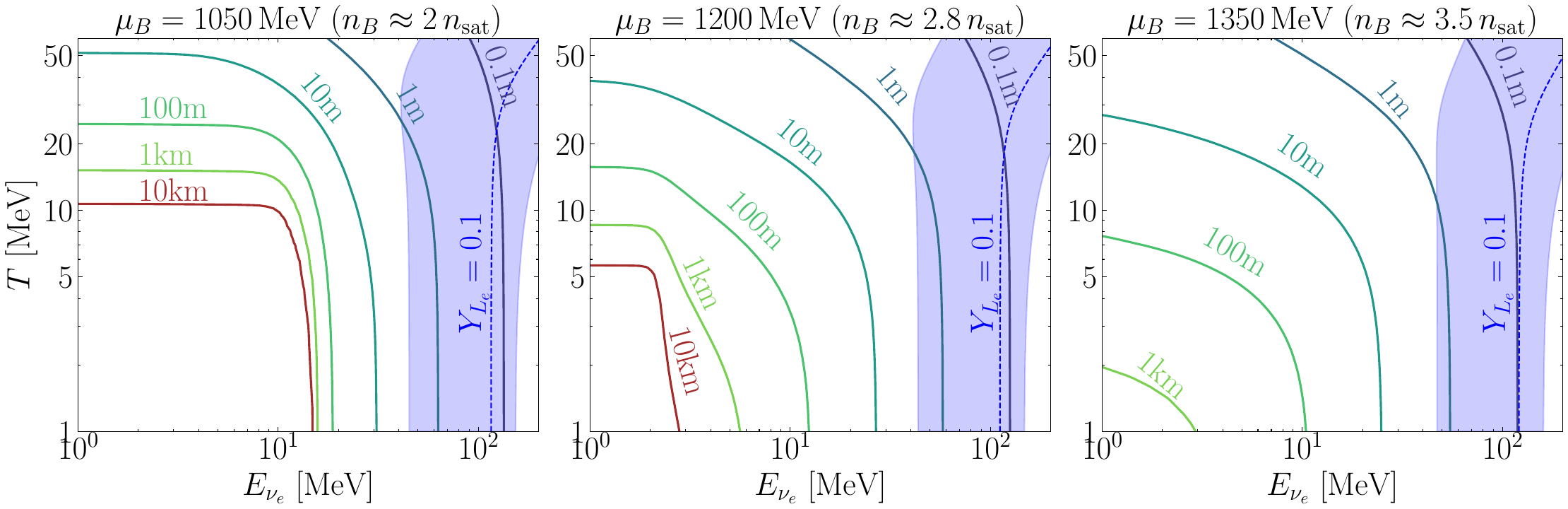}\\
\includegraphics[width=\linewidth]{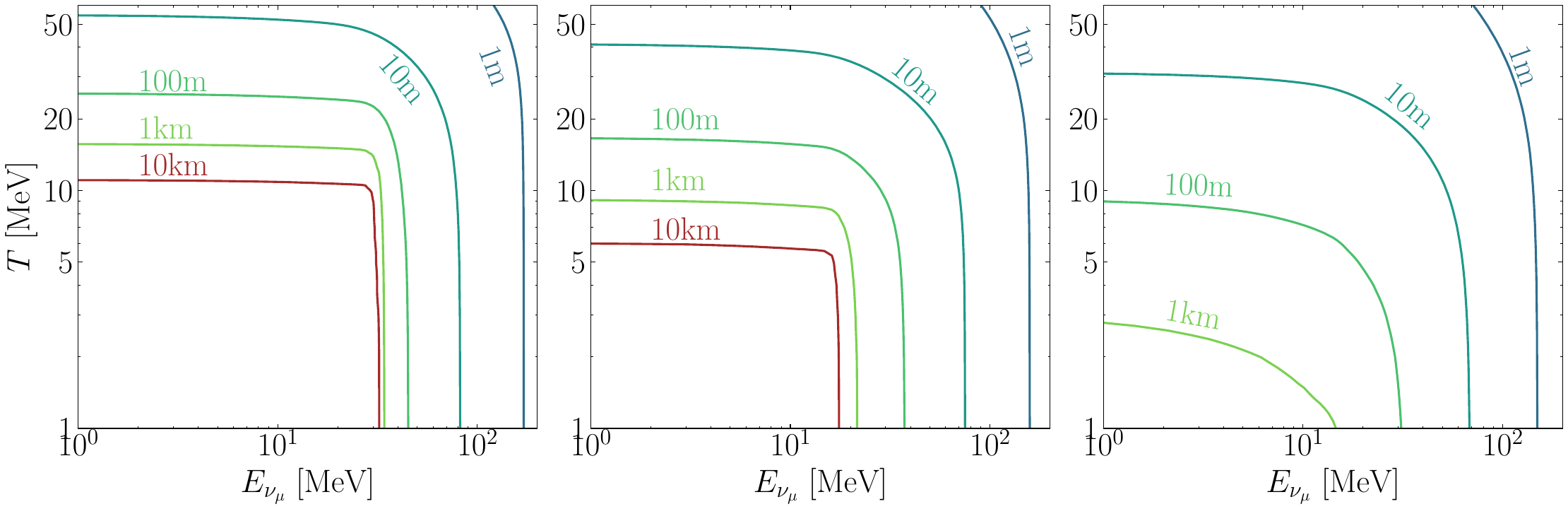}
    \caption{
Total absorption mean free path of electron neutrinos (top) and muon neutrinos (bottom) at three different baryon chemical potentials.  At the highest density (right-hand plot), there is a significant population of strange quarks.  The blue band in the plots for electron neutrinos depicts the 95\% confidence interval of a Fermi--Dirac distribution for neutrinos with the mean energy shown as a blue dashed line. For each temperature and density, the electron neutrino chemical potential which was used to calculate the blue band was determined by fixing the electron lepton fraction to $Y_{L_e}=0.1$, see text.
}
\label{fig:trapped}
\end{figure*}

\subsection{Neutrino mean free paths in the phase diagram}
\label{sec:thermal-phasediagram}

In Fig.~\ref{fig:pd} we show the density and temperature dependence of the absorption mean free path for electron neutrinos (upper panel) and muon neutrinos (lower panel) with the average thermal energy $E_\nu=3T$.

For electron neutrinos, we see that at temperatures above about $10\,\MeV$, the density dependence is mild, but at lower temperatures, there is a strong decrease of the mean free path with increasing temperature or density.
This reflects the results shown in Fig.~\ref{fig:density_dependence}, which as noted in Sec.~\ref{sec:density_dependence} are explained by the small momentum deficit for $d$-quark capture, which shrinks with increasing density, and can also be overcome by thermal blurring of the Fermi surfaces.
At densities above $3.2\,\nsat$ $s$-quark capture starts to become important: this explains the downward step in the $\la\gtrsim 100\,\text{m}$ contours at that density.

In the contour plot for muon neutrinos (lower panel of Fig.~\ref{fig:pd}), we see similar features. Here, there is a larger momentum deficit suppressing absorption by down quarks, which requires larger temperatures to activate the process, so at $T\lesssim 5\,\MeV$ the mean free path remains long at all densities up to $3.2\,\nsat$, at which point strange quarks appear, opening up the $s$-quark absorption channel, so the mean free path shrinks rapidly. Finally, we note again that Fig.~\ref{fig:pd} is for 2SC matter described by the NJL model, which does not describe hadronic phases that might be favored at low densities. Depending on the properties of the hadronic matter, astrophysical scenarios are therefore possible, for which neutrinos would leave the 2SC phase due to a long mean-free path at low temperatures and density, which, however, get then absorbed in the hadronic part of the system.

\subsection{Temperature dependence of the mean free path}\label{sec:T_dependence}

In Fig.~\ref{fig:T_dependence} we show the temperature dependence of the mean free path of a thermal ($E_\nu=3T$) electron neutrino (solid curves) or muon neutrino (dashed curves) at three fixed chemical potentials, which at zero temperature correspond to the densities $n_B\approx 2.0,\,2.8,\,3.5\,\nsat$.
We also include guide lines showing scaling as $T^{-2}$ (thin solid line)
and $T^{-3}$ (thin dotted line).

To understand the temperature dependence, we note that for strange quarks, the neutrino absorption process is kinematically allowed at zero temperature. Since we set $E_\nu=3T$, the mean free path is proportional to $T^{-2}$ \cite{Iwamoto:1982zz}. This explains the scaling of the $s$-quark capture mean free path (green lines) in the right panel.
The other two panels are for densities where the strange quark chemical potential is smaller than the strange mass, so the temperature dependence of $\la_{\nu+s}$ is dominated by the appearance of a thermal population of $s$ quarks.

For down quarks, the absorption process is not kinematically allowed at zero temperature, leading to a scaling proportional to $T^{-3}$ at temperatures that are large enough to overcome the momentum deficit. This is the behavior seen for $\la_{\nu+d}$ in all three panels. This modified scaling for the kinematically suppressed $d$-capture process compared to the allowed $s$-capture process has been discussed before in the context of flavor changing rates in the 2SC phase of the NJL model \cite{Alford:2024tyj,Alford:2025tbp}.

\subsection{Energy dependence of the mean free path}
\label{sec:neutrino-spectrum}

Previous sections have shown results for neutrinos with energy $E_\nu=3T$. We now allow the neutrino energy to vary independently of the temperature, 
and show how the mean free path depends on the neutrino energy.
Fig.~\ref{fig:trapped} shows how the total neutrino absorption mean free path depends on
neutrino energy $E_\nu$ and temperature $T$ at three different values of the baryon chemical potential. 

All the panels show a common pattern: the mean free path is longest for lower energy neutrinos, and at lower temperature. This can be understood via the kinematics of neutrino absorption processes: the mean free path depends on the available phase space, which grows as temperature rises (increasing the thermally blurred region around the Fermi surfaces) or as the incoming neutrino energy rises (opening up more phase space in the final state). The mean free path of the process is roughly determined by $\De Q_\text{th} \approx T + E_\nu$ (Eq.~\eqref{eq:Qth}), hence the contours approximately follow lines of constant $\De Q_\text{th}$, which on a log-log plot have the rounded-box shape seen in Fig.~\ref{fig:trapped}.

At the lowest baryon chemical potential $\mu_B=1050\,\MeV$ (left panels), corresponding roughly to a baryon density of $n_B\approx2\,\nsat$, the mean free path rises rapidly
as $T+E_\nu$ drops below the $d$-quark capture momentum deficit, which we see in the lower panel of Fig.~\ref{fig:density_dependence_down}.
is approximately $20\,\MeV$.
At higher densities (middle and right panels), the momentum deficit is smaller and easier to overcome via either thermal blurring or neutrino energy, leading to a more gradual variation of the mean free path.

To assess when the mean free path would be short enough to support a trapped gas of neutrinos, we overlay on the contour plots a blue dashed line and shaded region that shows, at each temperature, the mean neutrino energy and 5\% to 95\% percentile energy range of a Fermi--Dirac distribution
obtained by assuming a net lepton fraction $Y_{L_e}=(n_e+n_{\nu_e})/n_B=0.1$, a value at the upper end of the range expected in the central region in mergers \cite{Perego:2019adq,Pajkos:2024iry}. 
This requires\footnote{
Note that these neutrinos do not fulfill the 
trapped-neutrino beta equilibrium condition $\mu_d+\mu_{\nu_e}=\mu_u+\mu_e$, since we add them after
fixing the other particle fractions to obey
cold beta equilibrium, $\mu_d=\mu_u+\mu_e$.} $\mu_\nu\approx 100\,\MeV$, so
at temperatures up to $50\,\MeV$ the neutrinos form a degenerate gas, where the typical neutrino energy range is almost independent of the temperature.

We see that at all the densities shown this population of neutrinos would have 
mean free paths smaller than 10\,m, indicating that it is consistent to treat them as trapped and equilibrated within the star, and even within a $\sim 100\,\text{m}$ fluid element. This remains true at arbitrarily low temperatures, although at temperatures below a few MeV
non-degenerate neutrinos with $E_\nu\sim 3T$ would be free streaming, see Fig.~\ref{fig:T_dependence}. In the scenario of a neutron star merger simulation, it therefore depends on the history and surroundings of the fluid element under consideration whether a degenerate neutrino population with $\mu_\nu \simeq 100\,\MeV$ is a reasonable assumption or not.

Our results for the 2SC phase can be compared with the energy dependence of the neutrino mean-free path in nuclear matter, see e.g. Refs.~\cite{Ng:2023syk,Chiesa:2024lnu}.

\section{Conclusion}\label{sec:conclusion}

To understand the evolution and dynamics of compact stellar objects, it is necessary to investigate the interaction of neutrinos with strongly interacting matter. This includes  assessing under which conditions (densities, temperatures, and lepton fractions) the matter
can be treated as transparent or opaque for neutrinos. 

We used an NJL model to calculate neutrino absorption mean free paths in 2SC matter,
arising from the $d$-quark capture $\nu+d_b\to u_b+e^-/\mu^-$  and  $s$-quark capture $\nu+s_b\to u_b+e^-/\mu^-$ processes. We only included absorption by unpaired quarks: the contribution from paired quarks is negligible for temperatures well below the critical temperature of the 2SC condensate, $T_\text{2SC}\gtrsim 100\,\MeV$.

Our main findings are:
\begin{enumerate}
 \item In our model, the $d$-quark capture process $\nu_e+d_b\to u_b+e^-$ is suppressed by a momentum deficit that rapidly shrinks to a few MeV at densities above about $2.5\,\nsat$. The suppression can be overcome
by thermal blurring of the Fermi surfaces or the energy of the incoming neutrino, so the mean free path becomes
shorter at higher densities and temperatures, and for more energetic neutrinos. To capture this effect, one must integrate over the full phase space for the process: the Fermi surface approximation misses the opening of this process via thermal blurring.

 \item At densities above the onset of the strange quarks ($n_B \gtrsim 3.2\,\nsat$ in our NJL model), the $s$-quark capture process $\nu_e+s_b\to u_b+e^-$ is kinematically allowed and dominates the total mean free path at temperatures $T\lesssim 3\,\MeV$, where the $d$-quark capture process is suppressed.

 \item Over the temperature and density range that we studied, 2SC quark matter can support a population of trapped, equilibrated neutrinos. For example, in
 a fluid element with electron lepton fraction $Y_{L_e}=0.1$, equilibrated neutrinos would experience a typical mean free path of a few meters or less. 
\end{enumerate}

There are several directions in which this work can be developed.
\begin{enumerate}
    \item Our NJL model does not contain an interaction that would produce an isovector mean field (e.g., the $\rho$ meson). It would be interesting to explore models that contain such an interaction \cite{Chu:2016ixv,Chu:2016rpw,Liu:2019ntc,Liu:2023gmq} because it would introduce flavor-dependence in the energy shifts in the quark dispersion relations. The $d$-quark capture process could then be kinematically allowed
at some densities, changing some of the threshold-like behavior that we saw in the dependence of the mean free path on density, temperature, and neutrino energy.

\item  We calculated mean free paths for neutrinos, and it would be valuable to extend this
to antineutrinos. To see when this would be physically relevant, we refer to Fig.~\ref{fig:composition}, which shows that when the electron lepton fraction falls below about 0.08, there will be a significant antineutrino population at some densities. For muon neutrinos, this would happen when $Y_{L_\mu}$ falls below about 0.05, which is within the range typically found or assumed in merger simulations~\cite{Loffredo:2022prq,Gieg:2024jxs,Pajkos:2024iry,Ng:2024zve}.

\item We explored how thermal blurring can open up phase space for the (kinematically forbidden at $T=0$) $d$-quark capture process. However, there is another mechanism that can contribute to the rate: the in-medium modification of the quark excitations, arising from their strong interactions with the quark population. This is a quark matter analog of the ``modified Urca'' process in nuclear matter \cite{Knoll:1995nz,Raffelt:1998pa,Voskresensky:2001fd,vanDalen:2003zw,Alford:2024xfb}. Such contributions could be relevant for cooling of the 2SC phase in a system where the temperature is not high enough to overcome the momentum deficit, e.g.,~the late cooling stages of a hybrid proto neutron star.

\item 
While strong first-order phase transitions have been studied in gravitational wave signals \cite{Most:2018eaw,Bauswein:2018bma}, no simulation has examined the dynamic effects of chemical equilibration in quark matter. The required opacities (inverse mean free paths) for such a simulation are, to our knowledge, not publicly available in neutrino opacities databases like \texttt{NuLib}~\cite{OConnor:2014sgn,nulib}, \texttt{Weakhub}~\cite{Ng:2023syk} or \texttt{BNS\_NURATES}~\cite{Chiesa:2024lnu,Perego_BNS_NURATES_2025}. Our consistent microscopic calculation of mean free paths in quark matter is a step in this direction. 
\end{enumerate}

\section{Acknowledgements}
We thank Leonardo Chiesa,
Arus Harutyunyan, Melissa Mendes, Carlo Musolino, Armen Sedrakian, Jürgen Schaffner-Bielich, Andreas Schmitt, and Stefanos Tsiopelas for useful discussions and we thank Sophia Han and Harry Ho-Yin Ng for helpful comments on our manuscript. M.H. thanks the graduate school HGS-HIRe for Fair for financial support to visit Washington University in St.~Louis.
M.G.A. and L.B. are partly supported by the U.S. Department of Energy, Office of Science, Office of Nuclear Physics, under Award No.~\#DE-FG02-05ER41375. M.H. is supported by the GSI F\&E. M.B., H.G. and M.H. acknowledge support from the Deutsche Forschungsgemeinschaft (DFG, German Research Foundation) 
through the CRC-TR211 `Strong-interaction matter under extreme conditions' project number 315477589 – TRR 211. A.H.~acknowledges support by the U.S. Department of Energy, Office of Science, Office of Nuclear Physics, under Award No. DE-FG02-05ER41375. A.H.~furthermore acknowledges financial support by the UKRI under the Horizon Europe Guarantee project EP/Z000939/1.

\section*{Data Availability}
\noindent
The numerical data presented in all figures in this work are openly
available in the ancillary files of the corresponding arXiv
submission.
The equation of state and related microscopic quantities such as quark massses and diquark condensates of the NJL model presented in this work can be recalculated using the open-source NJL module \cite{hofmann_2026_18249033}, which was published as part of the most recent \textit{Calliope} version of the MUSES Calculation Engine~\cite{Jahan:2026hvs}.

\appendix

\section{Phase space for $d$-quark capture: comparison of NJL mean field with strong interaction corrections} \label{sec:app}

In previous work on weak interactions of unpaired quarks \cite{Baym:1975va,Iwamoto:1982zz,Steiner:2001rp,Colvero:2014zfa}, it is often assumed that strong interaction Fermi-liquid corrections to the quark dispersion relations will make $d$-quark capture kinematically allowed because the corrections alter the relationship between quark Fermi momentum and chemical potential in such a way that the triangle inequalities are all obeyed.
The correction for quark flavor $f$, assuming ultrarelativistic quarks, takes the form
\begin{equation}
    p_{F\!f}=\mu_f(1-\kappa_f) \ .
\end{equation}
If the correction is the same for up and down quarks, $\ka\equiv\ka_u=\ka_d$, then in beta equilibrium the $d$-quark capture momentum surplus is
\begin{align}
    p_F^u+p_F^e-p_F^d = (\mu_u-\mu_d)(1-\kappa)+\mu_e = \kappa(\mu_d-\mu_u) \ .
\end{align}

Since $\mu_d>\mu_u$ for nonzero quark masses, any $\kappa>0$ opens phase space for the $d$-quark capture process to happen. In Ref.~\cite{Wang:2006tg}, the shift $\kappa$ was calculated in perturbative QCD and in a two-flavor NJL model with a color-current vector interaction, both assuming an equal chemical potential $\mu$ for all quark flavors. In both cases, $\kappa(\mu)$ was found to increase monotonically with $\mu$ towards a positive value in the limit of large densities.

If the difference in chemical potential between up and down quarks is taken into account, then the correction is different for the two flavors.
In this case the momentum surplus is
\begin{align}
    p_F^u+p_F^e-p_F^d =  \kappa_d\mu_d-\kappa_u\mu_u.
\end{align}
Thus if $\kappa_d/ \kappa_u > \mu_u/\mu_d$, the $d$-quark capture process is allowed, while for $\kappa_d/ \kappa_u < \mu_u/\mu_d$ it is forbidden. Indeed, the latter is the case in our NJL model in the mean-field approximation.
This means, consistent with our discussion in the main text, that including interactions between quarks as in our mean-field NJL model has the opposite effect as to include strong interactions as in Ref.~\cite{Wang:2006tg}: the phase space for the $d$-quark capture process shrinks and is not enhanced. Whether $d$-quark capture is suppressed or not therefore depends on how the interactions between the unpaired quarks are modeled.

\bibliography{accepted.bib}
\bibliographystyle{apsrev4-1}

\end{document}